\documentclass[12pt]{article}
\usepackage{amsmath}

\input epsf

\DeclareSymbolFont{AMSb}{U}{msb}{m}{n}
\DeclareSymbolFontAlphabet{\Bbb}{AMSb}

\newcommand{\aD}{{\dot\alpha}}

\newcommand{\bD}{{\dot\beta}}
\newcommand{\U}{{U}}
\newcommand{\SU}{{SU}}
\newcommand{\SO}{{SO}}

\newcommand{\BL}{{\bf L}} 

\newcommand{\AdS}{AdS}

\begin{document}

\addtolength{\baselineskip}{2pt}
\thispagestyle{empty}

\begin{flushright}
UW/PT 98-18, DTP 98/74, SWAT-98/207\\
{\tt hep-th/9810243}\\
October 1998
\end{flushright}

\vspace{1cm}

\begin{center}
{\scshape\Large Multi-Instantons and Maldacena's Conjecture} 

\vspace{1cm}

{\scshape Nicholas Dorey$^{1,3}$, Timothy J.~Hollowood$^{2,3}$,
Valentin V.~Khoze$^4$,\\ Michael
P.~Mattis$^2$ and Stefan Vandoren$^3$}

\vspace{0.3cm}
$^1${\sl Physics Department, University of Washington,\\
Seattle, WA 98195, USA}\hspace{0.6cm} {\tt dorey@phys.washington.edu}\\

\vspace{0.2cm}
$^2${\sl Theoretical Division T-8, Los Alamos National Laboratory,\\
Los Alamos, NM 87545, USA} \\
{\tt mattis@lanl.gov}, {\tt pyth@schwinger.lanl.gov}\\

\vspace{0.2cm}
$^3${\sl Department of Physics, University of Wales Swansea,\\
Swansea, SA2 8PP, UK}\hspace{0.6cm} {\tt pysv@swan.ac.uk}\\

\vspace{0.2cm}
$^4${\sl Department of Physics, University of Durham,\\
Durham, DH1 3LE, UK}\hspace{0.6cm} {\tt valya.khoze@durham.ac.uk}\\

\vspace{1cm}

{\Large ABSTRACT}
\end{center}

\vspace{0.1cm}

\def\Z{{\cal Z}}
\def\N{{\cal N}}
\noindent We examine certain $n$-point functions
 $G_n$ in ${\cal N}=4$ supersymmetric $SU(N)$
gauge theory at the conformal point. 
In the large-$N$ limit, we are able to sum all
leading-order multi-instanton contributions exactly. We find
compelling evidence for Maldacena's conjecture: \hbox{(1) The}
large-$N$ $k$-instanton collective coordinate space has the
geometry of $AdS_5\times S^5\,$.
\hbox{(2) In} exact agreement with type IIB superstring calculations, at
the $k$-instanton level, 
$G_n =
\sqrt{N}\,g^8\,k^{n-7/2}e^{-8\pi^2k/g^2}\sum_{d|k}\,d^{-2}\cdot
F_n(x_1,\ldots,x_n)$,
where $F_n$ is identical to a convolution of $n$
bulk-to-boundary SUGRA propagators.

\newpage

\def\N{{\cal N}}
\def\det{{\rm det}}
\def\susic{supersymmetric}
\def\susy{supersymmetry}
\def\adss{AdS_5\times S^5}
\def\ads{AdS_5}
A remarkable conjecture by Maldacena \cite{MAL}  (see also
\cite{OTHER} and references therein) relates the large-$N$ limit
of 4-dimensional $\N=4$ \susic\ Yang-Mills theory (SYM) at the
conformal
point of vanishing VEVs, to the low-energy behavior of type IIB
superstrings on $\adss$. In this conjectured correspondence, the SYM
theory is defined on the boundary of the $\ads$; the generating
functionals of SYM correlation
functions are then equated to partition functions in the bulk with
specific boundary conditions. When the string coupling is small
and the radii of both the $\ads$ and
$S^5$ are large (namely small $g^2$ but large $g^2N$, where $g$ is the gauge theory
coupling) 
the string calculation can be approximated by classical supergravity,
whereas when $g^2N$ is small the SYM model is amenable to standard
perturbative and semiclassical analysis.

An important consequence of Maldacena's conjecture is that
Yang-Mills instantons should be related in a specific way to the
D-instantons of the IIB string theory on $\adss$.
In particular, Banks and Green \cite{BG}
showed that the $AdS_{5}\times S^{5}$ background solves the field equations 
of the same $SL(2,{\Bbb Z})$ invariant effective action which describes D-instanton 
corrections to the IIB theory in flat space. The proposed 
AdS/CFT correspondence then yields predictions for an infinite series of 
multi-instanton contributions in the ${\cal N}=4$ theory.
This relation was studied, at the
1-instanton level, in Ref.~\cite{BGKR} for gauge group $SU(2)$, and
in Ref.~\cite{DKMV} for general $SU(N)$. In this Letter, we extend the
1-instanton
results of \cite{DKMV} to arbitrary Yang-Mills multi-instantons. For
any
finite $N$,
the jump from single- to multi-instantons is always accompanied by
an enormous increase in calculational complexity, endemic to the ADHM
formalism. But in the large-$N$ limit there are compensating
simplifications, namely:

{\bf(i)} The integration over the $k$-instanton collective
coordinate space (``ADHM moduli space'') becomes dominated by $k$ single
instantons living in $k$ mutually commuting $SU(2)$ subgroups of
$SU(N)$. (In this one respect only, the problem becomes dilute-gas-like.)
This intuitively obvious remark, which we derive below from a
large-$N$ saddle-point treatment of the ADHM formalism,
 follows from statistical
considerations alone, and is independent of \susy. Additionally in
the large-$N$ limit, other less intuitive but equally remarkable
features appear, which depend crucially on the $\N=4$ \susy:

{\bf(ii)} To leading order in the large-$N$ saddle-point expansion,
the geometry of the
\hbox{$k$-instanton} moduli space is described by $(\adss)^k.$ Here the
$\ads$ factors are coordinatized by the positions and sizes
$(x_n^i,\rho_i)$
of the $k$ instantons, as has been noted previously \cite{BGKR,WIT112}. 
But the emergence
of the $S^5$ factors is a newly observed feature of large $N$ even at
the 1-instanton level; the
$(S^5)^k$ coordinates are the auxiliary antisymmetric-tensor variables
$\chi^i_{AB},$ $1\le i\le k,$ $1\le A,B\le4,$ used to bilinearize a
certain fermion quadrilinear term in the $\N=4$ instanton action.
One can check that the $SO(6)_{R}$ non-singlet operators which 
correspond to the Kaluza-Klein modes of the IIB  dilaton on $S^{5}$ couple to 
these angular degrees of freedom in the correct way \cite{BIG}.

{\bf(iii)} Integrations in the vicinity of
the large-$N$ saddle-point generate
an attractive
singular potential between the $k$ single instantons  that
draws them all to a common point, thereby reducing the moduli space
from $k$ copies to a single copy of $\adss.$ (In this respect the
problem is very $un$-dilute-gas-like.)

{\bf(iv)} At the $k$-instanton level, the leading-order small-fluctuations
effective Lagrangian expanded about this reduced moduli space turns
out to be identical to
10-dimensional \hbox{$\N=1$} \susic\ $SU(k)$ Yang-Mills theory on flat space,
dimensionally reduced
 to $0+0$ dimensions. The collective coordinate measure thus
factorizes into the measure on $\adss,$ times the partition function
$\Z_k$ for this $SU(k)$ theory. It is remarkable the latter factor is precisely
the partition function that describes D-instantons in {\it flat\/}
space \cite{GG}.

\def\G{{G}}
In what follows we will focus, for definiteness, on certain
gauge-invariant
$n$-point
functions $\G_n(x_1,\ldots,x_n)$ ($n=16,8$ or 4) 
of chiral superfield components that were defined in
Ref.~\cite{BGKR}. At the conformal point the SYM model has 16 exact,
unlifted, adjoint fermion zero modes (two supersymmetric plus two
superconformal modes, times four supersymmetries) that must be
saturated by explicit field insertions in the correlator. In $\G_{16}$
each such fermion is inserted once at a distinct space-time point, whereas in
$\G_8$ and $\G_4$ the relevant insertions consist of 8
fermion-bilinear and 4 fermion-quadrilinear pieces of bosonic
operators, respectively. 
The $G_n$ provide a way to compare 
the Yang-Mills and SUGRA sides of
Maldacena's conjecture.
On the SYM side, $G_n$ is given semiclassically by a
$k$-instanton collective coordinate integral; an independent
calculation is required for each $k$. In contrast, in the
D-instanton/SUGRA picture, $G_n$ has the effective-vertex structure of $n$
bulk-to-boundary propagators tied to a single point in $\adss$ that is
integrated over; the
$k$-instanton expansion simply consists of Taylor expanding the
overall coefficient, which is proportional to the modular form
$f_n(\tau,\bar\tau)$ \cite{GG}. As is emphasized in
Ref.~\cite{BGKR}, at the \hbox{1-instanton} level $G_n(x_1,\ldots,x_n)$ has
the identical propagator-like form  in both pictures. In fact, we show
below that this functional resemblance extends to all $k$,
due to properties {\bf(i)} and {\bf(iii)}. As a sharper test of Maldacena's
conjecture, we  calculate the leading semiclassical
 contribution to the coefficient of
$\G_n$ from each $k$-instanton sector in the SYM  model. 
Thanks to properties {\bf(i)}-{\bf(iv)}, the collective coordinate
 integrals can be carried 
out explicitly in the limit $N\rightarrow\infty.$ We find:
\begin{equation}
\G_n(x_1,\ldots,x_n)
\,{\Big|}_{k\hbox{-}\rm inst}\ =\  \sqrt{N}\,g^8\,k^{n-7/2}
e^{-8\pi^2k/g^2}\,
\sum_{d|k}\,d^{-2}\cdot{F}_n(x_1,\ldots,x_n)\ .
\label{ouranswer}\end{equation}
Here the $k$-independent function ${F}_n(x_1,\ldots,x_n)$ looks like a
convolution of $n$ bulk-to-boundary SUGRA propagators, as already noted,
while
the summation term over the positive integer divisors of $k$
comes from a recent evaluation of $\Z_k$ in
Refs.~\cite{GG,MNS,KNS}.

The SYM result \eqref{ouranswer} precisely matches the Taylor
coefficients of the corresponding D-instanton/SUGRA
 expression for $\G_n$ \cite{BGKR}. 
In our view, this highly nontrivial agreement, 
including the $x_i$ dependence---taken together with
the unexpected emergence of an $\adss$ moduli space in large-$N$ 
SYM---constitute convincing evidence in favor of Maldacena's conjecture. 

We should emphasize that the above comparison between the SYM and
SUGRA pictures can be quantitative if and only if there exists a
nonrenormalization theorem that allows one to relate the small $g^2N$
to the large $g^2N$ behavior of chiral correlators such as
 $G_n$, as has been suggested in
Refs.~\cite{BG,DKMV}. In the absence of such a theorem the best one
can hope for is that qualitative features of the agreement persist beyond leading order while 
the exact numerical factor in each instanton sector does not, in
analogy with the mismatch in the numerical prefactor between 
weak and strong coupling results for black-hole entropy \cite{GKP}. 
In our view, however, our present results provide strong
evidence in favor of such a 
nonrenormalization theorem for $G_n$, for the following reason.
Consider the planar diagram corrections to the leading
semiclassical (i.e., $g^2N\rightarrow0$)
result, Eq.~\eqref{ouranswer}. In principle, these would not only modify the
above result by an infinite series in $g^2N,$ but also, at each order
in this expansion, and independently for each value of
$k$, they would produce a different function
$F_n(x_1,\ldots,x_n).$ The fact that  the leading semiclassical 
form for $F_n$ that we obtain is not only $k$-independent, but
 already matches the D-instanton/SUGRA
prediction exactly, suggests that such planar diagram corrections actually
vanish.

This Letter is a shortened version of \cite{BIG}, which will contain
full calculational details, and will also show that
the connection between 
D-instanton  and SYM amplitudes is more general than the
large-$N$ context considered here. 
Maldacena's construction of the ${\cal N}=4$ theory starts from a configuration 
of $N$ D3 branes on ${\Bbb R}^{10}$. The world-volume theory of the 
D-instanton has a Higgs branch which is exactly the 
ADHM moduli space of $SU(N)$ \cite{MD}. 
In Ref.~\cite{BIG}, we will explain how the
supersymmetric multi-instanton measure 
obtained in Refs.~\cite{measure1,DHKM,KMS}, and used below, can
be deduced from string theory directly.

\def\M{{\cal M}}
\def\tr{{\rm tr}}
\def\trtwo{\tr^{}_2\,}
\def\Mbar{\bar{\cal M}}
\def\dalpha{{\dot\alpha}}
\def\dbeta{{\dot\beta}}
\def\dgamma{{\dot\gamma}}
\def\ddelta{{\dot\delta}}
\def\Skinst{S^k_{\rm inst}}
\def\dmuphys{d\mu^{k}_{\rm phys}}
In order to evaluate $G_n$ in the SYM picture, one inserts the $n$
appropriate gauge-invariant operators under the integration 
$\int\dmuphys\,\exp(-\Skinst)$
where $\Skinst$ is the $k$-instanton action and $\dmuphys$ is the
collective coordinate measure. These quantities are defined as
follows \cite{KMS}. The bosonic and fermionic collective coordinates
 live, respectively, in an
$(N+2k)\times2k$ complex matrix $a$, and in an $(N+2k)\times k$
Grassmann-valued complex matrix $\M^A,$ where the $SU(4)_R$ index
$A=1,2,3,4$ labels the \susy. In components:\footnote{We adopt the
conventions of Refs.~\cite{KMS,WB} throughout; see \cite{KMS} for
references to the early literature on ADHM. The indices
$u,v=1,\ldots,N$ are $SU(N)$ indices;
$\alpha,\dot\alpha,\hbox{etc.}=1,2$ are Weyl indices (traced over with
`$\trtwo$'); $i,j=1,\ldots,k$ ($k$ being the topological number) are
instanton indices (traced over with `tr$_k$'); and $m,n=1,2,3,4$ are
Euclidean Lorentz indices.
Collective coordinates integrals
are defined as in \cite{BIG,KMS}; for complex
bosonic integration our convention is 
$\int dz d\bar{z} \equiv \int d\, {\rm Re}(z)\, d \,{\rm Im}(z)$. }
\begin{equation}
a\ =\ 
\begin{pmatrix}
w_{ui\dalpha}\\ \big(a'_{\beta\dalpha}\big)_{li}
\end{pmatrix}
\ ,\qquad
\M^A \ =\ 
\begin{pmatrix}
\mu^A_{ui}\\ \big(\M^{\prime A}_\beta\big)_{li}
\end{pmatrix}
\label{adef}\end{equation}
where both $a'_m$ (defined by
$a'_{\beta\dalpha}=a'_m\sigma^m_{\beta\dalpha}$)
and $\M^{\prime A}_\beta$ are Hermitian $k\times k$ matrices. These
matrices are subject to polynomial ADHM constraints discussed shortly, as
well as to a $U(k)$ symmetry 
\begin{equation}
w_{iu\aD}\rightarrow w_{ju\aD}U_{ji}\ ,\quad
a'_{mij}\rightarrow U^{-1}_{ik}\,a'_{mkl}\,U^{}_{lj}\ ,\quad U\in U(k)\ ,
\label{E28}\end{equation}
that must be modded out when constructing physical quantities.
For future use we also introduce the boson bilinears
\begin{equation}
\big(W_{\ \bD}^\aD\big)_{ij}=\bar w_{iu}^\aD \,w_{ju\bD}\ ,\quad
W^0={\rm tr}_2\left(W\right),\quad W^c={\rm
tr}_2\left(\tau^cW\right), \ \ c=1,2,3\ .\label{E26}\end{equation}
We remind the reader that in the dilute instanton gas limit, the
diagonal elements of $W^0$ have the interpretation $W^0_{ii}=2\rho_i^2$
where $\rho_i$ is the scale-size of the  $i^{\rm th}$ instanton; likewise
$(-a'_m)^{}_{ii}$ is its 4-position.

In terms of these quantities, and 
in the absence of adjoint VEVs, the $\N=4$ $k$-instanton action is
given by \cite{DKMn4} 
\def\Skquad{S^k_{\rm quad}}
\begin{equation}
\Skinst\ =\ {8\pi^2k\over g^2}+\Skquad\ .
\label{E47}\end{equation}
Here $\Skquad$ is a particular fermion quadrilinear term, with one
fermion collective coordinate chosen from each of the four gaugino
sectors $A=1,2,3,4\,$:
\begin{equation}
\Skquad\ =\ {\pi^2\over g^2}\,\epsilon_{ABCD}\,{\rm tr}_k\left(\Lambda_{AB}
\BL^{-1}\Lambda_{CD}\right).
\label{E48}\end{equation}
The $k\times k$ anti-Hermitian fermion bilinear $\Lambda_{AB}$ is
defined by
\def\sqrtwo{\sqrt{2}\,}
\begin{equation}
\Lambda_{AB}\ =\ {1\over2\sqrtwo}
\big(\Mbar^A\M^B-\Mbar^B\M^A\big)
\label{Lambdadef}\end{equation}
whereas $\BL$ is a linear self-adjoint operator that maps the
$k^2$-dimensional space of such matrices
onto itself:
\def\hf{{\textstyle{1\over2}}}
\begin{equation}  
\BL\cdot\Omega\ =\ \hf\{\Omega,W^0\}-\hf{\rm tr}_2
\left([\bar a',\Omega]a'-\bar a'[a',\Omega]\right).
\label{E35}\end{equation}
As discussed in Ref.~\cite{DKMn4}, $\Skquad$ is a SUSY invariant which
lifts all the adjoint fermion modes except the 16 exact \susic\ and
superconformal modes. 

As in Refs.~\cite{KMS,DKMV}, it will prove useful to replace the
fermion quadrilinear $\Skquad$ with a fermion bilinear coupled to a
 set of auxiliary Gaussian variables. These take
the form of an anti-symmetric tensor $\chi_{AB}=-\chi_{BA}$ whose elements are
$k\times k$ matrices in instanton indices. The
expression we need is 
\begin{equation}\begin{split} 
&\exp\left(-\Skquad\right)\  =\\ 
&2^{9k^2}\,\pi^{-3k^2}\,
\big({\det}_{k^2}\BL\big)^{3}
  \int
d^{6k^2}\chi\exp\big[\,-\epsilon_{ABCD}{\rm tr}_k\left(\chi_{AB}\BL
\chi_{CD}\right)+4\pi ig^{-1}{\rm
tr}_k\left(\chi_{AB}\Lambda_{AB}\right)\,\big]\ .
\label{E53}\end{split}\end{equation} 
Requiring negative-definiteness of this Gaussian form is tantamount to
the specific reality condition
\begin{equation}
\hf
\epsilon^{}_{ABCD}\chi^{}_{CD}\ =\ \chi_{AB}^\dagger\ ,
\label{E54}\end{equation}
where the hermitian conjugation acts on instanton indices only. This
reality condition simply means that $\chi_{AB}$ transforms in the vector
representation of the $\SO(6)\cong\SU(4)$ R-symmetry. To make this manifest, we
can introduce a 6-vector of $k\times k$ matrices $(p^c,q^c)$, $c=1,2,3$
with
\begin{equation}
\chi_{AB}={1\over\sqrt8}\left(\eta^c_{AB}p^c+i\bar\eta^c_{AB}q^c\right),
\label{E54.1}\end{equation}
where the $\eta^c_{AB}$  ($\bar\eta^c_{AB}$) 
are the (anti-)self-dual 't Hooft  symbols
\cite{TH}. Equation \eqref{E54} then says that the
six matrices $(p^c,q^c)$ are hermitian.

Next we turn to the $k$-instanton $\N=4$ collective coordinate
measure,
which reads (see Ref.~\cite{DHKM} and Sec.~6 of Ref.~\cite{KMS}):
\def\wbar{\bar w}
\def\mubar{\bar\mu}
\def\abar{\bar a}
\begin{equation}
\begin{split}
\int\dmuphys\ &=\ {(C_1'')^k\over{\rm Vol}\big(U(k)\big)}\,
\int d^{2Nk}\wbar\,d^{2Nk}w\,d^{4k^2}a'\,\prod_{A=1,2,3,4}
d^{Nk}\mubar^A\, d^{Nk}\mu^A\,d^{2k^2}\M^{\prime A}
\\
&\times\ \big({\det}_{k^2}\BL\big)^{-3}\,
\Big[\prod_{c=1,2,3}\delta^{k^2}\big(\trtwo(\hf\tau^c\abar a)\big)\Big]
\Big[\prod_{A=1,2,3,4}\delta^{2k^2}\big(\Mbar^Aa+\abar\M^A\big)\Big]
\end{split}
\label{measuredef}
\end{equation}
The constant $C_1''$ is fixed at the 1-instanton level, by
comparing Eq.~\eqref{measuredef} with the standard 1-instanton 't Hooft-Bernard
measure \cite{TH, Bernard}; in the Pauli-Villars scheme one obtains
\begin{equation}
C_1''\ \underset{N\rightarrow\infty}{=}\ 2^8 \pi^{-6N} g^{4N}
\label{C1def}
\end{equation}
Note that the factors of $\det\BL$ cancel out between Eqs.~\eqref{E53}
and \eqref{measuredef}.
The matrix-valued arguments of the $\delta$-functions in Eq.~\eqref{measuredef}
are the usual bosonic and fermionic ADHM constraints, respectively.
An independent rederivation of this measure, directly from 
string theory, will be presented in Ref.~\cite{BIG}.

If, as in the present case, we intend to use our measure
to integrate only gauge-invariant quantities
(in particular, in the absence of adjoint VEVs), the expression 
\eqref{measuredef} can
be simplified by transforming to a smaller set of
gauge-invariant collective coordinates (i.e., variables without an
uncontracted `$u$' index). In the bosonic sector this means changing
variables from $\{w,\wbar\}$ to the $W$ variables of Eq.~\eqref{E26}. 
An interesting Jacobian identity is proved in Ref.~\cite{BIG}:
\begin{equation}
d^{2Nk}\wbar\,d^{2Nk}w\ =\ C_{N,k}\,\big(\det_{2k}W\big)^{N-2k}\,
d^{4k^2}W\ ,
\label{interesting}\end{equation}
where in the large-$N$ limit 
\begin{equation}
C_{N,k}\  \underset{N\rightarrow\infty}{=}\ 
2^{-k}\,e^{2kN}\,\big(\pi/N\big)^{2kN-2k^2}\ .
\label{E37}\end{equation}
We can already anticipate that as $N\rightarrow\infty,$ the Jacobian factor of
$(\det W)^N=\exp(N\,\tr\log W)$ will be amenable to a saddle-point treatment.
Another nice feature of this change of variables is that the bosonic
ADHM constraints in Eq.~\eqref{measuredef}, which are quadratic in the
$\{w,\wbar\}$ coordinates, become linear in terms of $W$; explicitly,
\def\sigmabar{\bar\sigma}
\def\etabar{\bar\eta}
\def\zetabar{\bar\zeta}
\def\mubar{\bar\mu}
\def\nubar{\bar\nu}
\begin{equation}
0\ =\ W^c \ + \ [\,a'_n\,,\,a'_m\,]\,\trtwo
(\tau^c\sigmabar^{nm})\ =\ 
 W^c \ - \ i \ [\,a'_n\,,\,a'_m\,]\,\etabar^c_{nm}\ .
\label{E27}\end{equation}
We therefore write $d^{4k^2}W\,=\,2^{-2k^2-k}\,d^{k^2}W^0\prod_{c=1,2,3}
d^{k^2}W^c$ and use Eq.~\eqref{E27} to eliminate  $W^c$ from the measure.

Likewise in the fermion sector we change variables from 
$\{\mu,\mubar\}$ to $\{\zeta,\zetabar,\nu,\nubar\}$ defined by
\begin{equation}
\mu_{iu}^A=w_{uj\aD}\,\zeta^{\aD A}_{ji}+\nu_{iu}^A\ ,\qquad
\bar\mu_{iu}^A=\bar\zeta_{\aD ij}^A\,\bar w_{ju}^\aD+\bar\nu_{iu}^A\ ,
\label{E44}\end{equation}
where the $\nu$ modes form a basis for the $\perp$-space of $w\,$:
\begin{equation}
0\ = \ \bar w_{iu}^\aD\nu_{ju}^A\ =\  \bar\nu_{iu}^Aw^{}_{ju\aD}\ ,
\label{E45}\end{equation}
In these variables the fermionic ADHM constraints in
Eq.~\eqref{measuredef} have the gauge-invariant form
\begin{equation}
0\ =\ \zetabar^A\,W+W\,\zeta^A+[\M^{\prime A},a']
\label{E46}\end{equation}
which can
 be used to eliminate 
 $\bar\zeta^A$ in favor of  $\zeta^A$ and ${\cal M}^{\prime A}$;
doing so  gives a factor which precisely cancels
the Jacobian for the change of variables \eqref{E44}.

 Notice that
 $\nu$ and $\nubar$ are absent from  the constraint   
\eqref{E46}, and only appear
in the coupling $\tr_k^{}\,\chi_{AB}\Lambda_{AB}$ in Eq.~\eqref{E53}. 
We eliminate them from the measure, as follows.
First, we decompose
 $\Lambda_{AB}=\hat\Lambda_{AB}+\tilde\Lambda_{AB}$,
where 
\begin{equation}
(\hat\Lambda_{AB})_{ij}={1\over2\sqrt2}\left(\bar\nu_{iu}^A
\nu_{ju}^B-
\bar\nu_{iu}^B\nu_{ju}^A\right)\ ,
\label{E51}\end{equation}
and
\begin{equation}
\tilde\Lambda_{AB}={1\over2\sqrt2}\left(\ 
\bar\zeta^A W\zeta^{B}-\bar\zeta^B W\zeta^{A}
+
[{\cal M}^{\prime A}\,,\,{\cal M}^{\prime B}]\ \right)\ .
\label{E52}\end{equation}
Second, we calculate
\begin{equation}
\int d^{4k(N-2k)}\nu\,d^{4k(N-2k)}\bar\nu\,\exp\left({4\pi i\over
g}\,{\rm tr}_k(\chi_{AB}\hat\Lambda_{AB})\right)=
\left(8\pi^2\over
g^2\right)^{2k(N-2k)} \left({\rm det}^{}_{4k}\chi\right)^{N-2k}.
\label{E55}\end{equation}
Note the similarity of this result to the Jacobian in
Eq.~\eqref{interesting}; it too will contribute to the saddle-point
equations in the large-$N$ limit. The third and final contribution to
these equations will be the Gaussian term in $\chi$ in Eq.~\eqref{E53},
once one rescales
 $\chi_{AB}\rightarrow\sqrt N\chi_{AB}$ so that $N$ factors out in
front. Combining the above manipulations, one now finds for the
gauge-invariant measure:
\begin{equation}\begin{split}
\int\dmuphys\,e^{-\Skinst}\ =\ &
 {g^{8k^2}N^{k^2}e^{-8\pi^2k/g^2}
\over2^{2k^2-6k-8}\,\pi^{13k^2}\,{\rm Vol}(\U(k))}\int d^{k^2}W^0\,d^{4k^2}a'\,
d^{6k^2}\chi\,\prod_{A=1,2,3,4}d^{2k^2}\M^{\prime A}\,
d^{2k^2}\zeta^A
\\
&\times\ \left({\det^{}_{2k} W}{\det^{}_{4k}\chi}\right)^{-2k}\,
\exp\big[4\pi i g^{-1}\sqrt{N}\,{\rm tr}_k(\chi_{AB}\tilde \Lambda_{AB})
\ -\ NS^k_{\rm eff}\big]
\label{E57.2}
\end{split}\end{equation}
where the ``effective $k$-instanton action'' is the sum of the three
aforementioned terms (plus a constant piece):
\begin{equation} S^k_{\rm
eff}=-2k\left(1+3\log2\right)-\tr_{2k}\log W-\tr_{4k}\log
\chi +\epsilon_{ABCD}\,{\rm tr}_k\left(\chi_{AB}\,
\BL\,\chi_{CD}\right)\ .
\label{E57}\end{equation}

\def\Skeff{S^k_{\rm eff}}
We now formally pass to the limit $N\rightarrow\infty,$ and
stationarize $\Skeff$ with respect to the
$11k^2$ variables $a'_m$, $W^0$ and $\chi_{AB}$. 
The general solution to the
coupled saddle-point equations reads (with no sum on $i$):
\begin{equation}
(W_{\ \bD}^\aD)_{ij}^{}=\rho_i^2\delta_{ij}\delta^\aD_{\ \bD}\ ,\qquad
\left(\chi^{}_{AB}\right)_{ij}={1\over\rho_i}X_{AB}^i\delta_{ij}\ ,\qquad
\left(a'_m\right)_{ij}=-x_m^i\delta_{ij}\ ,
\label{E61}\end{equation}
up to an adjoint action on all these quantities
by the $\U(k)/\U(1)_{\scriptscriptstyle\rm diag}^k$
 coset of the 
$\U(k)$ symmetry \eqref{E28}. For each value of the instanton index
$i=1,\ldots,k$, the quantity $X_{AB}^i$ satisfies
\begin{equation}
\epsilon^{}_{ABCD}X_{AB}^iX_{CD}^i=1 \qquad\hbox{(no sum on $i$)}\ .
\label{E62}\end{equation}
Equivalently, 
in terms of the $\SO(6)$ vector variables \eqref{E54.1},
Eqs.~\eqref{E61}-\eqref{E62} imply that
the six matrices $(p^c,q^c)$ are diagonal with elements
\begin{equation}
q^c_iq^c_i+p^c_ip^c_i={1\over\rho_i^2}\qquad(\text{no sum on }i).
\end{equation}
In other words,  up to  $U(k)/\U(1)^k$, 
our $k$-instanton solution is parameterized by $k$ points in
the 10-dimensional space
$AdS_5\times S^5$, where, for each instanton, $(x^i_m,\rho_i)$
 are the coordinates in $AdS_5$
and $(p_i^c,q_i^c)$ is a point on $S^5$. 
One is ineluctably led to the
conclusion that at large $N$ the $k$-instanton measure 
in ${\cal N}=4$ SYM theory at the conformal point
is related to a configuration of $k$
D-instantons in string/SUGRA theory!
Note also that the effective action \eqref{E57} evaluated on the
solution is zero.

This charge-$k$ solution
 also has a very simple interpretation from the point of
view of the gauge theory. As per our initial intuition, it
consists of $k$
individual $\SU(2)$ instantons, with  positions $x_m^i$ and scale sizes
$\rho_i$, all embedded in mutually commuting $\SU(2)$ subgroups of the $SU(N)$
gauge group. Indeed the  embedding of the $i^{\rm th}$ instanton is
determined by the vectors $w_{iu\aD}$, from which one forms the three
$SU(2)$ generators
\begin{equation}
\left(t^c\right)_{uv}={1\over\rho_i^2}w_{iu\aD}\left(\tau^c\right)^\aD_{\
\bD}\bar w_{iv}^\bD\qquad \hbox{(no sum on $i$)}\ .
\end{equation}
The fact that, at the saddle point,
 $(W_{\ \bD}^\aD)^{}_{ij}=0$ for $i\neq j$, is precisely the statement
that the two $\SU(2)$ embeddings commute. 

The next step in the saddle-point expansion is
to integrate the Gaussian fluctuations around
the $k$-instanton solution. Here
 we encounter an important phenomenon
which actually alters the naive $N$ counting of the measure.
 When one analyses the quadratic fluctuations around the solution
\eqref{E61}, one finds, generically, $10k$
zero modes corresponding to the parameters of the solution and $k(k-1)$
zero modes corresponding to the action of the $\U(k)/\U(1)^k$ 
symmetry \eqref{E28}.
However, it can be shown that
when pairs of instantons are at the same point in $AdS_5\times S^5$
there are additional zeroes of the quadratic operator; the
correspondingly singular small-fluctuations determinant then acts
 as an attractive singular potential on the instanton moduli space.
As detailed in Ref.~\cite{BIG}, it turns out
that such special solutions are more dominant in the large $N$ limit
and, indeed, the leading-order $N$ dependence comes from the
completely degenerate solution when {\it all\/} the instantons sit at
the same point in $AdS_5\times S^5$: 
\begin{equation}
x_m^i\equiv x_m\ ,\quad \rho_i\equiv\rho\ ,\quad X_{AB}^i\equiv
X_{AB}\ .
\label{degen}\end{equation}
 Notice that this special solution is completely invariant
under $\U(k)$.
 
Around the special solution, the fluctuations fall into three
sets. First, there are $10$ true zero modes which correspond to the
position of the $k$-instanton
``bound state'' in  $AdS_5\times S^5$. Second, there
are $k^2$ fluctuations of the form 
\begin{equation} 
\varphi=2\rho\epsilon_{ABCD}X_{AB}\delta\chi_{CD}+{1\over2\rho^2}\delta W^0,
\end{equation}
which have a nonzero quadratic coefficient in the small-fluctuations expansion.
 The remaining $10k^2-10$
fluctuations first appear beyond quadratic order and they correspond to 
the traceless parts
$\hat\chi_{AB}$, or $\{\hat p^c,\hat q^c\}$, and $\hat a'_m$ of the ten $k\times k$ matrices 
$\chi_{AB}$ and
$a'_m$. The bosonic part of the measure splits as follows:
\begin{equation}\begin{split}
d^{k^2}W^0\,d^{4k^2}a'\,
d^{6k^2}\chi\ &=\ k^52^{-15k^2+2k+5}\rho^{-2k-5}\, d^4x\, d\rho\,
d\Omega_5\\
&\times\ d^{k^2}\varphi\, d^{4(k^2-1)}\hat a'\, d^{3(k^2-1)}\hat
p^c\, d^{3(k^2-1)}\hat q^c\,
\end{split}\end{equation} 
where $d\Omega_5$ is the volume element for the solid angle of $S^5$
parameterized by $X_{AB}$ and the integrals over the traceless matrices
are defined with respect to a basis normalized to ${\rm
tr}_k(T^aT^b)=\tfrac12\delta^{ab}$ with $1\le a,b\le k^2-1$.

It turns out that in the expansion of the effective action around the
special solution, there is no cubic coupling involving 
three variations of $\hat a'_m$ and $\hat\chi_{AB}$. This 
means that at leading order in $N$ 
\begin{equation}
\varphi\sim{\cal O}\left(N^{-1/2}\right),
\qquad\hat a'_m\sim{\cal O}\left(N^{-1/4}\right),\qquad
\hat\chi_{AB}\sim{\cal O}\left(N^{-1/4}\right).
\end{equation}
The Gaussian fluctuations $\varphi$ can be integrated out explicitly,
leaving an additional contribution to the quartic coupling of the
fluctuations $\hat a'_m$ and $\hat\chi_{AB}$. To complete the
expansion, the fermion 
terms in \eqref{E57} involve 
the traceless parts $\hat\zeta^{\aD A}$ and $\hat{\cal
M}^{\prime A}_\alpha$ coupled to $\hat a'_m$ and $\hat\chi_{AB}$ and the 
fermionic measure factorizes as
\begin{equation} 
d^{2k^2}\zeta^A\, d^{2k^2}{\cal M}^{\prime A}\ =\
k^{-2}\,2^{2k^2-2k-8}d^2\xi^A\,d^2\bar\eta^A\,d^{2(k^2-1)}\hat\zeta^A\,
d^{2(k^2-1)}\hat{\cal M}^{\prime A},
\end{equation} 
where $\xi^A_\alpha$ and $\bar\eta^{\aD A}$ are the 16
supersymmetric and superconformal modes.  
  
Remarkably, in the large-$N$ limit,
the leading-order terms of the effective action
around the special solution, with the quadratic fluctuations
$\varphi$ integrated out, precisely assemble themselves into the
dimensional reduction from ten to zero of ${\cal
N}=1$ supersymmetric Yang-Mills with gauge group $\SU(k)$ in {\it
flat\/} space. The $\SU(k)$ adjoint-valued ten-dimensional gauge field and 
Majorana-Weyl fermion are defined in terms of the fluctuations:
\begin{equation}
A_\mu=N^{1/4}\left(\rho^{-1}\hat a'_m,\rho\hat p^c,\rho\hat
q^c\right),\qquad
\Psi=\ 
\Big({\pi\over2g}\Big)^{1/2}\,N^{1/8}\left(\rho^{-1/2}\hat{\cal
M}^{\prime A}_\alpha,\rho^{1/2}\hat\zeta^{\aD A}\right).
\end{equation}
The action for the dimensionally reduced gauge theory is 
\begin{equation}
S(A_\mu,\Psi)=-{1\over2}{\rm tr}_k\,\left[A_\mu,A_\nu\right]^2\ 
+\ {\rm
tr}_k\left(\bar\Psi\Gamma_\mu\left[A_\mu,\Psi\right]\right).
\end{equation}
It might have been anticipated that the action would depend on $X_{AB}$
in some way. Actually it does, but only through 
the representation of the ten-dimensional gamma matrices $\Gamma_\mu$
\cite{BIG}.

We conclude that the effective gauge-invariant
measure for $k$ instantons in
the large-$N$ limit
 factorizes into a 1-instanton-like piece, for the position
of the bound state in $AdS_5\times S^5$ and the $16$ 
supersymmetric and superconformal modes, times the partition function 
${\cal Z}_k$ of the
dimensionally-reduced ${\cal N}=1$ supersymmetric $\SU(k)$  gauge theory
in flat space:
\begin{equation}\begin{split}
\int\dmuphys\,e^{-\Skinst}\ \underset{N\rightarrow\infty}{=}\ &
{\sqrt Ng^8e^{-8\pi^2k/g^2}\over k^32^{k^2/2 -k/2
+24}\,\pi^{-k^2/2+13}{\rm Vol}(\U(k))}\\
&\times\ \int
\rho^{-5}d\rho\,d^4x\, d\Omega_5\prod_{A=1,2,3,4}d^2\xi^A 
d^2\bar\eta^A\cdot{\cal Z}_k\ ,
\label{hello}\end{split}\end{equation}
with
\begin{equation}
{\cal Z}_k={1\over(2\pi)^{5(k^2-1)}}\int dA_\mu\,
 d\Psi\,e^{-S(A_\mu,\Psi)},
\end{equation}
and where the appropriate group volume factor is 
\begin{equation}
\qquad{\rm Vol}(\U(k))={2^k\pi^{k(k+1)/2}\over\prod_{i=1}^{k-1}i!}\ .
\end{equation}
The normalization of the partition function ${\cal Z}_k$ has been
chosen to agree with \cite{KNS}. When integrating
$\SU(k)$ singlet quantities, which is relevant for the comparison to
string theory, the factor ${\cal Z}_k$ is simply a constant. It was
known \cite{GG,MNS} that this constant is proportional to
$\sum_{d\vert k}\, d^{-2}$ (a sum over the integer divisors of $k$);
however, the constant of proportionality has only been recently
settled \cite{KNS}:
\begin{equation}
{\cal Z}_k\ =\ {2^{k(k+1)/2-1}\,\pi^{(k-1)/2}\over\sqrt k\prod_{i=1}^{k-1}i!}
\sum_{d\vert k}{1\over d^2}\ .
\end{equation}
Using this expression, our final expression for the measure on
gauge invariant and $\SU(k)$ singlet operators in the large $N$
limit is
\begin{equation}
\int\dmuphys\,e^{-\Skinst}\underset{N\rightarrow\infty}{=}
{\sqrt Ng^8k^{-7/2}\over2^{25}\pi^{27/2}}
\,e^{-8\pi^2k/g^2}
\sum_{d\vert k}{1\over d^2}\cdot\int\rho^{-5}
d^4x\,d\rho\, d\Omega_5\prod_{A=1,2,3,4}d^2\xi^A 
d^2\bar\eta^A.
\label{hellone}\end{equation}

Finally, we can  use our measure to calculate the correlation functions
$G_n(x_1,\ldots,x_n)$. This entails inserting into Eq.~\eqref{hellone}
the appropriate product of gauge-invariant composite chiral operators
$\Phi_1(x_1)\times\cdots\times\Phi_n(x_n)$, which together contain
the requisite 16 exact fermion modes, and
approximating each $\Phi_j$ by its $k$-instanton saddle-point value
$\Phi_j^{(k)}.$ Explicit forms for the $\Phi_j$ are given in
Ref.~\cite{BGKR}, and need not be repeated here. However, we can make
the following observation.  Since,
at leading order in $N$, the $k$  instantons sit
at the same point in $\AdS_5\times S^5$, albeit in mutually commuting
$SU(2)$ subgroups of $SU(N)$, it follows that 
$\Phi_j^{(k)}$ is simply
proportional to its single-instanton counterpart:
$\Phi_j^{(k)}=k\Phi_j^{(1)}$.
Therefore $G_n$ scales like $k^n$. Furthermore, the spatial function
$F_n(x_1,\ldots,x_n)$ is precisely the same as in the 1-instanton
sector, where its reinterpretation as a convolution of SUGRA
propagators, as dictated by Maldacena's conjecture, 
has already been emphasized in Ref.~\cite{BGKR}.
The final result of our Yang-Mills instanton calculation is summarized in
Eq.~\eqref{ouranswer}, where we absorb the constant factor $2^{-25}\pi^{-27/2}$
into $F_n$.

$$\scriptscriptstyle{************************}$$
The UKSQCD Collaboration thanks Ed Corrigan and Ivo Sachs for
discussions. TJH would like to thank the T-8 group
at Los Alamos for their hospitality.
ND, VVK and MPM acknowledge a NATO Collaborative Research Grant,
ND, TJH and VVK acknowledge the TMR network grant FMRX-CT96-0012
and SV acknowledges a PPARC Fellowship for support.

\end{document}